\documentclass[aps,preprint,a4paper,floatfix]{revtex4-1}

\usepackage{graphicx,color}
\usepackage{amsmath,amssymb,amsfonts,amsthm,amscd,bm}
\usepackage{booktabs}
\usepackage{cancel}

\usepackage{amsmath}
\usepackage{amssymb}
\usepackage{amsfonts}
\usepackage{amsthm}
\usepackage{amscd}
\usepackage{bm}
\usepackage{graphicx}
\usepackage{booktabs}
\usepackage{listings}
\usepackage{xcolor,colortbl}
\usepackage{float}

\def\tilde{\widetilde}

\def\hat{\widehat}
\def\*{\star}
\def\[{\left[}
\def\]{\right]}
\def\({\left(}      

\def\){\right)}

\def\frac#1#2{\dfrac{#1}{#2}}
\def\inv#1{\dfrac{1}{#1}}

\def\2pi{\hbog{$2\pi i$}}

\def\dsl{\raise.15ex\hbox{/}\kern-.57em\partial}
\def\Dsl{\,\raise.15ex\hbox{/}\mkern-.13.5mu D}
\def\Om{\Omega}

\def\2pi{\hbox{$2\pi i$}}

\def\dsl{\raise.15ex\hbox{/}\kern-.57em\partial}
\def\Dsl{\,\raise.15ex\hbox{/}\mkern-.13.5mu D}

\def\barray{\begin{eqnarray}}
\def\earray{\end{eqnarray}}
\def\beq{\begin{equation}}
\def\eeq{\end{equation}}

\def\AA{\leavevmode\setbox0=\hbox{h}
\dimen0=\ht0 \advance\dimen0 by-1ex\rlap{\raise.67\dimen0\hbox{\char'27}}A}

\def\rhohat{\hat{\rho}}
\def\phat{\hat{p}}
\def\ptilde{\tilde{p}}

\def\a{\mathfrak{a}}
\def\adot{\dot{\a}}

\def\ti{t_i}
\def\tf{t_f}
\def\rsi{r_{s,{\rm inf}}}
\def\R0{R_0}
\def\T0{T_0}
\def\t0{t_0}
\def\Om{\Omega_{\rm rad}}

\def\bigAbsl{\Big |}
\def\bigAbsr{\Bigr |}

\begin{document}

\bibliographystyle{plainnat}

\title{{\Large {\bf Early inflation  based on an aboriginal black hole}
 }} 

\author{
 Andr\'e  LeClair\footnote{andre.leclair@gmail.com}
}
\affiliation{Cornell University, Physics Department, Ithaca, NY 14850}

\begin{abstract}
\noindent
\noindent

It has been recently proposed that the interior of a black hole is vacuum energy,   and this offers a possible explanation for the dark energy of the current observable universe.     In this article we explore  the idea that black holes are a source of vacuum energy in  the early universe  as well,  and could possibly explain early inflation.    Our scenario 
pre-supposes a single  aboriginal black hole  as the source of vacuum energy that drives inflation and  eventually evaporates due to Hawking radiation.   
In a simplified  model  we can  roughly estimate $\ti$ and $\tf$ which are the times when inflation began and ended respectively.    Our  model leads to 
$\ti \approx  10^{-45}$ sec,     $\tf \approx 10^{-43} $ sec and 
  the number of e-foldings  during inflation  to be  about $70$.

\end{abstract}

\maketitle

\section{Introduction}


Theories of inflation provide  compelling explanations for several puzzles in early universe cosmology \cite{Guth,Linde,Steinhardt1}.  The theory essentially explains the initial conditions  for the Big Bang.    
For instance,  it  can solve the horizon problem,    and thereby  explain why the universe appears to be isotropic,  why the cosmic microwave background (CMB)  radiation is smooothy distributed,  and why the universe is flat,    
which are major successes.     It also potentially explains the large scale structure of the universe as arising from quantum fluctuations.      
The subject has evolved considerably since the original papers.     Soon after Guth's initial paper,  Linde,   and Albrecht and Steinhardt,  proposed new inflation  to  address the
``graceful exit" problem.    Over the last few decades,  physicists have proposed chaotic,  eternal, hybrid,  cyclic, and string based models of inflation.   For reviews see
\cite{Guth2,Steinhardt2,StringCosmo}.    As emphasized by Guth \cite{Guth2},    all these variations of inflation theory are plausible and  essentially share the same successes of the original inflationary scenario, so that some version of inflation very likely occurred.

Generally speaking,   models of inflation invoke a  false vacuum for a hypothetical  ``inflaton"  field, and the vacuum energy of  this field drives the exponential expansion.   New inflation involves a ``slow roll" of the inflaton to its true vacuum.      Specific models require some amount of fine tuning of the inflaton potential.    The origin of the inflaton and its potential remains unknown.
    For this reason,  it is worthwhile to explore other possible sources of vacuum energy that could also lead to an exponential expansion of the scale of the universe at very early times,   and the aim  of this paper is to consider one such possibility.

Here  we suggest a possible model of early inflation based on a single primitive black hole in the early universe,  which we will refer to as ``aboriginal".    The early universe was small and dense  enough that it could have behaved as a black hole,  at least  momentarily.   It was argued in \cite{LeClairBH} that the interior of a black hole is actually vacuum energy,  with  no singularity at the origin.       For the current universe,   one can view black holes as pockets of vacuum energy that contribute to the observed dark energy density,  leading to an accelerated expansion,   namely a ``late"  inflation.        
We roughly estimated the density of this dark energy and found  very good  agreement with the currently observed value of  approimately  $ 10^{-30}   {\rm g}/{\rm cm}^3$  \cite{CC1,CC2}.          This proposed source of dark energy is not a conventional cosmological constant since it changes over time,  depending on the relative density of black holes in the universe.        

 It is natural then to investigate  whether this source of vacuum energy due to black holes  could account for early inflation,  since the latter requires some form of vacuum energy. 
    At this early time,  the black holes that we have argued \cite{LeClairBH}  contribute to the observed dark energy of late inflation have not yet been formed.   
This idea  requires a specific scenario since black holes are static solutions to Einstein's equations,   whereas inflation is a time dependent expansion.     

 Let us summarize our possible scenario as follows.    At some time $\ti$,   the Hawking temperature of the aboriginal  black hole was such that it equaled the temperature of the universe.   It was thus in a state of thermal equilibrium 
and did not Hawking radiate.    At this moment it served as a source of vacuum energy and led to an  exponential expansion of the universe.      Eventually,  at some time $\tf$,   the temperature of the universe decreased enough that the black hole was out of equilibrium with the universe,  since the black hole  was hotter.    It  completely evaporates,   inflation ends, and the early hot radiation  dominated era began.    Our model has no inflaton field,  no  false vacuum, and  essentially no free parameters. 
This scenario  is clearly speculative,  and completely different than previous  inflationary models.   We present  it  here since our estimates of $\ti$,  $\tf$ and the number of e-foldings during inflation are very  reasonable.     Below we estimate $\ti \approx  10^{-45}$ sec,  $\tf \approx 10^{-43} $ sec and 
  the number of e-foldings  to be  about $70$.      This value for $\ti$ is close to the Planck time,  but as we will see this appears to be coincidental,  since,  in addition to $\hbar$,  our expression below  for $\ti$  actually involves 
  the current Hubble constant $H_0$,  Boltzmann's constant,  the current CMB temperature $T_0$,  and $\Om$.               

 In the literature one typically finds quotations of  the values $\ti \approx 10^{-36}$ sec and $\tf \approx 10^{-33}$ to $10^{-32}$  sec.
  This appears to go back to the initial idea of Guth that the source of vacuum energy could have been the spontaneous symmetry breaking at the GUT scale,   which leads to this value of $\ti$.    However,  the actual  value of $\ti$ is not a strong prediction  of the standard inflationary models;   what is more important is the number of e-foldings,   which were predicted long ago to be in the range $60-70$.

In the next section we present a summary of the particular  exact solutions to Einstein's equations found in \cite{LeClairBH} that are relevant to this article (many more can be found 
there).   Then in the subsequent section we analyze the above scenario with the goal of estimating  $\ti$,  $\tf$,  and the number of e-foldings.    
In equations we  mainly set $c=1$,  but restore it in places in order to calculate certain quantities.


\section{The interior of a static black hole as vacuum energy}

In this section we summarize the analysis in  \cite{LeClairBH}  that led to the proposal that the interior of a black hole is vacuum energy.   We introduced a length scale $r_0$ and assumed the stress-energy tensor had the following property:

\beq
\label{Tzero}
T_{\mu \nu} = 0  ~~{\rm for}  ~ r> r_0, ~~~~~T_{\mu\nu} \neq 0 ~~{\rm for} ~ r \leq r_0.
\eeq
Thus,  for $r>r_0$ the solution must be equal to the Schwarzschild solution with total mass $M$ inside $r_0$.     
In finding solutions,  one needs to match  the  Schwarzschild solution at 
$r=r_0$.

 The general static metric is defined by the line element 
 \beq
 \label{metric}
 ds^2 = g_{\mu\nu} dx^\mu dx^\nu = - e^{2a} \, dt^2 + e^{2b}\,  dr^2 + r^2 ( d\theta^2 + \sin^2 \theta  \,d\varphi^2 ).
 \eeq
We take the following form for the stress-energy tensor:
\barray
T_{tt} &=& \rho \, e^{2a}, ~~~~~T_{rr} = p\,  e^{2b}
\\
T_{\theta \theta} &=& \tfrac{1}{\sin^2 \theta} T_{\varphi \varphi} = p_\theta \, r^2 
\earray
where $\rho$ is the energy density and $p , p_\theta$ are pressures,  where $p$ is a radial pressure,  and $p_\theta$ an orbital one. 
 
It is convenient to rescale $\rho$ and  the  pressures  as follows
\beq
\label{hats} 
\rhohat = 8 \pi G\, \rho, ~~~~ \phat = 8\pi G\, p, ~~~~~ \ptilde = 8 \pi G\, p_\theta 
\eeq
such that $\rhohat, \phat$ and $\ptilde$ all have dimensions of inverse length squared,  actually inverse time squared if  one reintroduces $c$.  
The Einstein equations now read
\barray
\label{E1}
\frac{2b'}{r}  + \inv{r^2} \( e^{2b}-1 \) &=& \rhohat \,e^{2b} 
\\
\label{E2}
\frac{2a'}{r} - \inv{r^2} \( e^{2b} -1 \) &=& \phat \, e^{2b} 
\\
\label{E3}
a'' + a'^2 - a' b' +(a'-b')/r &=& \ptilde \,e^{2b} 
\earray
where $a'(r)$ denotes $da/dr$, etc.     
The second order equation \eqref{E3} can be replaced    with a first order equation by differentiating \eqref{E2} and using \eqref{E1}.   One obtains 
the  complicated equation:
\beq
\label{E3prime}
\ptilde = (4(r^2 \rhohat -1) )^{-1}  \[ r^2 \phat^2 (2rb'-1) + 2 \phat \( r^2 \rhohat(1+rb') + rb' -2 \) 
+ r \( 2 \rhohat ( b' + r^2 \phat' ) -2 \phat' -r\rhohat^2 \) \]
\eeq
which turned  out to be  useful. 
Thus one first solves the first order equations  \eqref{E1} and \eqref{E2}, and then $\ptilde$ is determined by \eqref{E3prime}.

\label{zerop}

Since $\rhohat$ and $\phat$  have units of inverse length squared,  let us define the scale $\ell$ 
\beq
\label{elldef} 
\rhohat \equiv  1/ \ell^2  , ~~~~ \phat =w \rhohat
\eeq
where $\ell$  has units of length,  which we  take to be constant in $r$,  and $w$ is an equation of state parameter.     
The solution to \eqref{E1} and \eqref{E2}  is
\barray
\label{Sol1}
e^{2b} &=& \(  1 - \frac{r^2}{3 \ell^2} - \frac{\beta \ell}{r} \)^{-1} 
\\
\label{Sol2} 
e^{2a} &=&  C^2 ~ \( \frac{\ell}{r} \)\,  \prod_{i=1}^3 \bigAbsl 1-  \frac{r}{x_i} \bigAbsr^{\nu_i}
\earray
where $\beta$ and $C^2$ are constants of integration, 
\beq
\label{nuw} 
\nu_i = \frac{ 1 + w\,  x_i^2/\ell^2}{1- x_i^2/ \ell^2}, 
\eeq
and ${x_i}$ are the three roots to the cubic polynomial equation 
\beq
\label{xroots} 
3 \beta - 3 (x/\ell) + (x/\ell)^3 = 0.
\eeq
Matching the above solution to the Schwarzschild solution at $r=r_0$,   one obtains
\barray
\label{rsrobeta}
\frac{r_s}{r_0} &=& \frac{r_0^2}{3 \ell^2} + \frac{\beta \ell}{r_0} 
\\
\label{C2o}
C^2 &=& \frac{r_0}{\ell} \( 1- \frac{r_s}{r_0} \) \, \prod_{i=1}^3 \bigAbsl  1- \frac{r_0}{x_i}  \bigAbsr^{-\nu_i} 
\earray
where $r_s = 2 G M$ is the Schwarzschild radius with  $M$  the total mass inside  $ r\leq r_0$.

\def\bigAbsl{\Big |}
\def\bigAbsr{\Bigr |}

We insist on non-singular solutions in $e^{2b}$;   thus we set $\beta =0$.    Then,   $\{ x_i \} /\ell= \{0, \pm \sqrt{3} \}$.  One sees that the 
$x_i =0$ root   cancels  the $1/r$ factor in  \eqref{Sol2}.     (See \cite{LeClairBH} for more details.).       
One obtains
\beq
\label{e2anu}
e^{2b} = \(  1 - \frac{r^2}{3 \ell^2}  \)^{-1} , ~~~~~
e^{2a} = C^2  ~ \bigAbsl 1-\frac{r^2}{3\ell^2} \bigAbsr^{\nu},   ~~~~~~~~~~~~~\nu = -(1+ 3 w)/2 .
\eeq
Matching to the Schwarzschild solution at $r=r_0$,  one finds
\beq
\label{wMatch}
\frac{r_s}{r_0} = \frac{r_0^2}{3 \ell^2} ,~~~~~~C^2 =  \( 1- \frac{r_s}{r_0} \)  \bigAbsl  1- \frac{r_s}{r_0} \bigAbsr^{- \nu}  .
\eeq
The constraint $r_s/r_0 = r_0^2/3 \ell^2$  just implies $\tfrac{4}{3} \pi r_0^3 \,  \rho = M$.     

We wish to describe the interior of a black hole,  which corresponds to $r_0 \to r_s$,   approaching the limit from above.   
If $w>-1$,   then $C^2 \to 0$,   whereas if $w <-1$,  $C^2 \to \infty$.    Thus the only physically sensible choice is $w=-1$,  where $C^2 =1$.      
Remarkably,   from \eqref{E3prime} one obtains the non-trivial result  from this  complicated expression  that $\ptilde = -1/\ell^2$.   Thus 
\beq
\label{cconstant}
\{ \rhohat, \phat , \ptilde \} = \{ 1, -1, -1 \}/\ell^2 ,
\eeq
which is entirely consistent with vacuum energy $T_{\mu\nu} = - \rho \, g_{\mu\nu}$.   
    Inside the black hole,  or more precisely  inside the event horizon  $r<r_s$,  the solution is particularly simple:
\beq
\label{BHsolution}
e^{2a}   =   1-\frac{r^2}{3 \ell^2}, ~~~~~~ e^{2b} = \(1- \frac{r^2}{3 \ell^2} \)^{-1}  ~~~~~~( r<r_s ).
\eeq
   The original black hole singularity at $r=0$  of the Schwarzschild solution has disappeared.  Inside the black hole,  
the energy density and pressures are interpreted as vacuum energy due to \eqref{cconstant}.

\def\xvec{{\bf x}}

\section{An Early inflation scenario}

Let $\t0$ denote the present time,  and $\ti$ and $\tf$  the times when inflation started and ended respectively in the early universe.     Our aim is to estimate
$\ti$ and $\tf$ based on the idea that the early universe was dense and small enough that it behaved like a black hole.    This scenario was sketched out in the Introduction.   
The arguments of the last section 
suggest  that  this  hypothetical aboriginal black hole was a source of vacuum energy.  

The FLRW metric is 
\beq
\label{RWmetric}
ds^2 = -dt^2 + \a(t)^2  \, d\xvec \cdot d\xvec 
\eeq
where $\xvec$ are the three spatial coordinates.   
The Friedman equations can be reduced to  
\barray
\label{Friedman}
\( \frac{\adot}{\a} \)^2  &=&  \frac{8 \pi G}{3} \, \rho
\\
\dot{\rho} &=& -3 \( \frac{\adot}{\a} \) (\rho + p )
\earray
where $\adot$ denotes a time derivative.  

In order to make any progress analyzing our proposed scenario,   we need to make some assumptions concerning the temperature of the universe  as a function of time.  Earlier than  about  $1$ second,  the universe  is believed to have been radiation dominated,   where  $p = \rho/3$.   
Solving the Friedman equations with only radiation as a source,   one has  the well-known result 
\beq
\label{radiation} 
\frac{\rho_{\rm rad}}{\rho_c} = \frac{\Om}{\a(t)^4}
\eeq
where $\rho_c = 3 H_0^2/8 \pi G$ is the critical energy density with $H_0 =  \adot/\a$   the Hubble constant at the present time $t_0$:   
$H_0   \approx 2 \times 10^{-18} ~ {\rm sec}^{-1}$.    It is known that $\Om \approx 0.00005$ \cite{CC2}.      
This leads to the known result
\beq
\label{asol} 
\a (t) = \sqrt{2 H_0   \, \Om^{1/2} \, \, t}.
\eeq
The radiation dominated era is the the most reliable way to track the temperature of the universe as a function of time.   Since
$\rho_{\rm rad} \propto T^4$,   one has 
\beq
\label{Tt}
T(t) =  \frac{T_0}{\a (t)}
\eeq
where $T_0$ is the current CMB  temperature.

Recall that  in our proposed  scenario,  the aboriginal black hole was in  thermal equilibrium  with the universe and thus not   losing energy  to Hawking radiation.  
Since we are extrapolating the temperature based on the radiation dominated area,   this would seem to imply we are assuming the black hole coexists with some radiation. However this point is not entirely clear since we cannot present detailed mechanisms for  the evaporation of the black hole into radiation;   thus let us leave this issue aside and proceed.     
 Reintroducing $c$ for calculational purposes, 
the Hawking temperature is \cite{Bekenstein,Hawking} 
\beq
\label{Hawk0}
T_H =   \frac{\hbar \, c }{4 \pi\, k_B\, r_s }
\eeq
where $r_s = 2 M G/c^2$ is the Schwarzschild radius and $k_B$ Boltzmann's constant.      We impose  that the black hole was in thermal equilibrium with the universe:
\beq
\label{Hawk}
T(t_i) = \frac{T_0}{\a ( \ti )} = T_H  (\ti) = \frac{\hbar \, c }{4 \pi\, k_B\, \rsi  }
\eeq
where $T_0 = 2.7 K$ is the current temperature of the CMB,   and    
 $\rsi \approx  c\,  \ti$  is the approximate radius of the black hole at $\ti$.       The above equation has one independent variable $\ti$,  and the solution is  
\beq
\label{tinf}
\ti = \( \frac{\hbar}{k_B T_0} \)^2  \, \frac{H_0}{8 \pi^2} \,\,  \Om^{1/2}  ~ \approx~ 1.4 \times 10^{-45} ~ {\rm sec} 
\eeq
which is an interesting new time scale for cosmology.  

 One also has 
\beq
\label{rsivalue}
\rsi \approx  c \, \ti \approx  4.3 \times 10^{-35} ~{\rm cm}.
\eeq
which recall is the approximate  size of the black hole when inflation began.   
The mass of the aboriginal black hole at $\ti$ was thus  about $3 \times 10^{-7}$ grams,   which is approximately one hundredth of the Planck mass. 
The temperature at $\ti$ was about $4 \times 10^{32} \, K$,   thus the Hawking temperature  was  very high,  as expected.       

Interestingly, the above numbers are all around the Planck scales of time, space,  mass, and temperature,  except perhaps for the mass of the aboriginal black hole.     This appears coincidental since  the explicit expression \eqref{tinf} 
for $\ti$ 
shows that it has little to do with the Planck time $t_p = \sqrt{\hbar G/c^5} = 5.4 \times 10^{-44}$ sec.    

The entropy of the aboriginal black small is extremely low at the start of inflation.   The Bekenstein-Hawking entropy is 
\beq
\label{BH2}
S = \frac{k_B}{4 \ell_p^2}  \, \, A 
\eeq
where  $\ell_p = \sqrt{\hbar G/c^3}$ is the Planck length and $A= 4 \pi  \rsi^2$.   
Based on the above estimate \eqref{rsivalue} for $\rsi$,  one finds
\beq
\label{Entropy}
S/{k_B}  = \log \Omega  \approx 0.002 ~~     \Longrightarrow ~~ \Omega \approx 1.002;.
\eeq
Thus $\Omega$ is approximately $1$ which is the lowest it could possibly be.     This possibly explains why the universe should have begun  in a state of very low entropy
in order for the current universe to have very high entropy.

We now turn to estimating $\tf$,   the time when inflation ends.    In our scenario,  as the universe expands due to the vacuum energy of the black hole,   its temperature decreases and the black hole and the radiation are no longer in equilibrium,  so that the black hole completely  evaporates due to Hawking radiation and inflation ends.    
We estimate $\tf$ as follows.    It is straightforward  to show that   
\beq
\label{rhohatt}
\rhohat (t) = 8 \pi G \,  \rho=  \frac{3}{4 t^2} 
\eeq
in the radiation dominated era.    
We {\it define}  $\tf$ as the time when the scale factor $\a$ matches that of the radiation  dominated era.   It is beyond our scope to describe the precise processes involved in this short period,   since some of the radiation presumably results from the evaporation of the black hole.      

During inflation,    the black hole is treated as vacuum energy based on the arguments of the last section.    Solving the Friedman equations during inflation,  one finds the usual exponential expansion due to  vacuum energy,
\beq
\label{expansion}
\a_{\rm inf}  (t) = \exp \( \sqrt{\tfrac{\rhohat_i}{3}}\, t \)
\eeq
where $\a_{\rm inf}$ is the scale factor during inflation and 
\beq
\label{rhohatt}
\rhohat_i  =  \rhohat (\ti)
\eeq
is a  very large constant.   
The above criterion for $\tf$ leads to 
\beq
\label{tfeq}
\rsi \,\,  e^{\tf/2\ti}  = \a(\tf) \, R_0
\eeq
where $\a (t)$ is given by \eqref{asol} and $R_0$ is the current radius of the universe at $t_0$, roughly   
$R_0 \approx 9 \times 10^{28} ~ {\rm cm}$.
The equation \eqref{tfeq} is a single equation for the unknown $\tf$.   It is a transcendental equation that can solved explicitly in terms of the Lambert W-function.   
Since  we are only interested in a single $\tf$ with all other parameters in the equation given,  we just solve  it numerically and obtain:
\beq
\label{tfvalue}
\tf  \approx 2 \times 10^{-43} ~ {\rm sec}.
\eeq

One promising feature of the above calculation is that the number of e-foldings it predicts is 
\beq
\label{efold}
{\rm e-foldings} =  \frac{\tf}{2 \ti} \approx 74
\eeq
which is consistent with the original predictions of inflationary models.  (See the review \cite{Guth2} for instance.)     The latter  predictions were based not so much on details of the inflaton potential but rather on the number of 
e-foldings necessary to sufficiently smooth out the universe and solve the horizon problem, etc.

\section{Concluding Remarks}

We have considered  an inflationary scenario wherein the source of vacuum energy is a single original black hole rather than an inflaton field.       Our estimate of the number of e-foldings during inflation is consistent with  the original predictions of inflationary theory.      Although our estimates of various parameters seem promising,  and justify our presentation of the idea,      further scrutiny of the  consistency of this proposal is  certainly warranted.

Our estimate \eqref{tinf}  for the time of the start of inflation  $\ti$  is close to the Planck time  scale,  but this appears  coincidental.  The mass of the aboriginal black hole is sub-Planckian,  namely about one hundredth of the Planck mass,  which is more consistent with the string  scale, and to a slightly lesser extent,  the GUT scale, for reasons we cannot explain.                  One should thus wonder  whether quantum gravity effects would affect our proposal.   We have in fact  used the Hawking temperature in obtaining \eqref{tinf},   but this is not based on  a complete theory of quantum gravity,  so this remains an open question.

\section{Acknowledgements}

We wish to thank the support of INFN and SISSA in Trieste,  Italy   where most of this work was carried out in June 2019.  

\bigskip 
\noindent{{\bf Note added:}   After completing this article we were informed of previous works on non-singular black holes with  negative pressure from a somewhat different perspective.  
In particular we wish to cite  the works 
\cite{Hayward,Mottola,Ramy1,Ramy2} and references therein.        


\end{document}